\documentclass[runningheads]{llncs}
\usepackage{graphicx}
\begin{document}
\title{Assisting International Migrants with Everyday Information Seeking: From the Providers’ Lens}
%
%

\author{Yongle Zhang \orcidID{0000-0002-4887-946X} \and
Ge Gao \orcidID{0000-0003-2733-2681}}

\authorrunning{Y. Zhang and G. Gao}

\institute{University of Maryland, College Park, United States\\
\email{\{yongle,gegao\}@umd.edu}}
%
%
%
\maketitle              
\begin{abstract}
International migrants face difficulties obtaining information for a quality life and well-being in the host country. Prior research indicates that international migrants often seek information from their co-national cohort or contacts from the same country. The downside of this practice, however, is that people can end up clustering in a small-world environment, hindering the information seekers' social adaptation in the long run. In the current research, we investigated the ongoing practices and future opportunities to connect international migrants with others beyond their co-national contacts. Our work zooms in on the providers' perspectives, which complements previous studies that pay exclusive attention to the information seekers. Specifically, we conducted in-depth interviews with 21 participants assisting the needs of informational migrants in the United States. Some of these people are fellow migrants from a different home country than the information seeker, whereas the rest are domestic residents. Our data revealed how these participants dealt with language barriers, overcame knowledge disparities, and calibrated their effort commitment as information providers. Based on these findings, we discuss directions for future information and communication technologies (ICT) design that can facilitate international migrants' daily information seeking by accounting for the provider's needs and concerns.

\keywords{information provider \and   international migrant \and  language barriers \and knowledge gap \and commitment \and}
\end{abstract}
\section{Introduction}
International migrants require information for daily tasks in a new society, such as accessing healthcare and transportation, understanding local laws, and finding jobs or housing. Access to adequate information helps migrants overcome everyday challenges (e.g.,\cite{caidi2008information,smith2011review}) and alleviate their acculturative stress \cite{ward2001psychology}. 

However, obtaining the necessary information can be challenging for international migrants living in their host country. For example, many people lack the basic knowledge about where to begin their information search \cite{chung2015exploratory,yeh2003international}. Language barriers also pose an issue for people whose native language differs from the host country’s majority \cite{mao2015investigating,suh2019had}. Official resources for international migrants often prioritize academic or professional development \cite{liao2007information,liu1997information}, which fail to cover other aspects of a person's everyday information needs.

Scholars have recognized social contacts of international migrants as critical information sources in their everyday information seeking \cite{hendrickson2011analysis,mao2015investigating,oh2016newcomers}. Migrants sharing the same country of origin (i.e., co-nationals) often play a pivotal role \cite{alzougool2013finding,mehra2007glocal,oh2016newcomers,tsai2021help}, due to shared language, culture, and migration experiences \cite{komito2011migrants,kudo2003intercultural}. These co-nationals, while usually motivated to assist, can sometimes limit migrants' exposure to diverse local perspectives due to their network's size and homogeneity \cite{jeong2004unbreakable,rui2015social,ward1991impact}. 

Other two roles also exist in an international migrant’s network: fellow migrants from different home countries, and domestic individuals who have grown up and lived in the migrant’s host country (i.e., domestic residents). These groups offer the potential to introduce diverse resources \cite{forbush2016social,williams2011can}, yet migrants often do not fully utilize them \cite{oh2014information}. While several initiatives have been developed to bridge the gap \cite{hirsch2004speakeasy,schwarz2015help}, their sustainability remains a challenge. Many of these projects heavily emphasize the viewpoints of information seekers, potentially overlooking the challenges faced by the information providers \cite{seguin2022co}.

Our work incorporated the missing perspective by conducting in-depth interviews with 21 information providers, including 11 fellow migrants and 10 domestic residents in the United States. Our study offers two key contributions: 1) theoretical contributions by examining the practice of assisting international migrants with daily information seeking, emphasizing providers’ perspectives—a group largely overlooked by prior research, and 2) practical contributions by highlighting barriers and design recommendations shared by our participants. We compared our work with previous research and discussed how our work contributes to Information Studies and Human-Computer Interaction for improving international migrants’ quality of life.

The current research builds upon existing efforts in information studies to enhance the daily information of international migrants. In particular, we aim to incorporate the perspective of information providers. We ask the following research questions (RQ):

RQ1: What challenges, if any, do information providers face in their current practice of assisting international migrants' daily information seeking?

RQ2: How can future ICTs be designed to help the information providers overcome those challenges? 

\section{Related Work}

We review the existing literature on two facets of international migrants' daily information seeking: the challenges they face and the potential information sources they utilize. 

\subsection{Daily Information Seeking for International Migrants}

Information seeking is essential for international migrants as they navigate various aspects of life in the new environment \cite{chung2015exploratory,komito2011migrants,oh2016newcomers,sin2011international}. In our study, international migrants are defined as individuals temporarily living outside their country of birth or citizenship \cite{caidi2010information}, such as international students and migrant workers. They often encountered challenges in everyday life \cite{caidi2008information,caidi2010information}. For instance, people need to find suitable housing accommodations (e.g., \cite{mao2015investigating,mehra2007glocal}), manage finances (e.g., \cite{yoon2017international}), and access healthcare services (e.g., \cite{oh2016newcomers}). These obstacles impede migrants from effectively navigating their new environment \cite{mao2015investigating,mehra2007glocal,sin2011international,thomas2010transnational,wang2015immigration,yoon2017international}, highlighting the importance of better comprehending and enhancing international migrants’ information-seeking practices.  

One line of this research indicates that international migrants often lack awareness of local information resources in the host country. For example, Mehra and Bilal interviewed various groups of Asian international students in the United States \cite{mehra2007glocal}. They found that locating essential daily information was both cognitively taxing and time-consuming for participants. Similarly, the diary studies conducted by Yoon and Chung suggested that migrant participants frequently did not know where to source information needed for their lives abroad \cite{yoon2017international}. Unfamilarity with local resources, and ICT differences between migrants’ home and host country contribute to these navigational barriers \cite{tang2018new,wyche2012we}. 

Another challenge faced by many migrants is gathering and understanding information in a host country where their native language is not predominantly spoken \cite{berendt2009user}. For instance, Rózsa et al. observed how Hungarian students performed online information seeking in English \cite{rozsa2015online}. They found that participants often crafted vague or incomplete English queries, yielding low-quality search results. Such language barriers have been frequently reported by participants in studies examining newcomer adaptation \cite{gao2022taking,jeong2004unbreakable,kudo2003intercultural}. While providing resources in multiple languages can be a solution, existing infrastructures in host countries typically cater to the language(s) spoken by the majority, leaving smaller linguistic groups underserved \cite{blasi2021linguistic,cobo2013exploration,han2015mental,pavlenko2008m}. 

International migrants often underutilize resources offered by professionals and officials \cite{jackson2005incoming,liao2007information,liu2009chinese,mao2015investigating,mehra2007glocal,yoon2017international}. For instance, public libraries with trained staff ready to help international visitors are frequently underused \cite{sin2015demographic}. Some researchers attribute this underutilization of institutional resources to the limited range of services offered. In other words, the information available at public libraries mainly pertains to an international migrant's academic performance or specific domain achievements (e.g., \cite{liao2007information,yi2007international}), which does not encompass the full array of everyday information needs in society \cite{jeong2004unbreakable,liu1997information,sakurai2010building}. Considering the above complexities, it is important to examine ways that support international migrants’ information-seeking practices. 

\subsection{Social Contacts as Information Providers}

Scholars in information studies have examined how people interact with their information sources. For example, in Wilson's framework \cite{wilson1981user,wilson1994information}, people's choices of information sources are influenced by variables such as the characteristics of the sources (e.g., accessibility). Johnson's model examines the characteristics and utility of information channels selected and used by seekers \cite{johnson2012health}. It discusses how a person's antecedents' factors (e.g., demographics and needs) shaped the uses of sources. In the context of everyday life information seeking (ELIS), Savolainen emphasizes individuals' tendencies to utilize informal daily sources, particularly outside of academic or professional settings \cite{savolainen2005everyday}. Researchers have applied the concept of ELIS to investigate migrants' information sources and found that their social networks are a primary source that offers valuable information across their settlement process \cite{jeong2004unbreakable,oh2016newcomers,sin2015demographic}. These social contacts fall into three categories: co-national migrants, fellow migrants, and domestic residents \cite{bochner1977friendship}.
\subsubsection{Co-nationals as information providers.} Co-nationals, international migrants from the same country of origin, serve as vital information providers for their peers. Their shared linguistic, cultural, and experiential backgrounds make co-nationals invaluable sources of both emotional and informational support \cite{kudo2003intercultural,maundeni2001role}. There's ample evidence highlighting the importance of co-nationals in helping migrants navigate their daily lives (e.g., \cite{oh2016newcomers,oh2014information}). For example, a study by Mao \cite{mao2015investigating} involving Chinese migrants in Canada indicated a preference for fellow Chinese when seeking everyday information and services. Newly arrived migrants often experience high stress and anxiety, making them inclined to seek comfort in the familiar environment provided by their co-national community \cite{hurh1990religious}.

Despite the benefits, previous research has cautioned people to beaware of the risks caused by over-dependence on their co-nationals as information sources \cite{lee2011internet,tsai2021help}. The size of an international migrant’s network in the host country is relatively small. Repeated information exchange within this small network can not only trap the person in an echo chamber \cite{chatman1991life,jeong2004unbreakable,ward1991impact}; it may also negatively affect people’s adaptation process in the long run \cite{chang2017digital,chang2020way}.
 
\subsubsection{Potential information providers beyond a migrant's co-nationals} Besides co-nationals, a migrant's social contacts include domestic residents and fellow migrants from other countries. These groups are recognized for their potential to introduce new perspectives and resources \cite{hamid2015information，smith2011review}. For example, Alho interviewed international migrants with experience of seeking employment in the host country \cite{alho2020you}. Participants acknowledged that expanding their social network beyond co-nationals could allow them to better comprehend local job market and personnel selection process. Surveys conducted by Forbush et al. suggested that international students with more diverse networks reported better social adaptation than those with fewer domestic connections \cite{forbush2016social}. However, empirical studies have shown that international migrants' network often contains a high proportion of co-nationals \cite{alberts2005there,bochner1977friendship,yan2013chinese,ye2006traditional}, limiting their access to diverse information. This trend may not change as migrants’ social interactions do not necessarily become more intertwined with locals over time \cite{gomes2014home，yuan2013understanding}.

Several research projects have leveraged ICT to connect newcomers with these prospective information providers and utilize them for daily information seeking. For example, Hirsch and Liu designed Speakeasy, an integrated web and telephone-based service connecting newcomers with knowledgeable local volunteers in the Boston area \cite{hirsch2004speakeasy}. These providers register their contact details and language skills, allowing the system to match them with information seekers who require assistance. Similarly, Brown and colleagues designed Rivtran, a system that enables the asynchronous exchange of audio and textual messages about everyday information seeking between immigrants and mentors assigned by the local community center \cite{brown2016designing}.

Although these systems provide the technical framework for international migrants to connect with local contacts \cite{brown2016designing,gatteschi2012match,hirsch2004speakeasy,ngan2016refugees}, field testing data indicate that finding available and qualified volunteers to satisfy the migrants’ information needs is often not easy \cite{schwarz2015help}. Many initiatives are small-scale and short-term practices, presenting sustainability concerns. In successful information-seeking practices, both information seekers and providers play crucial roles in ensuring effective communication and knowledge exchange \cite{case2010model,kuhlthau1991inside,robson2013building}. For instance, the Information Seeking and Communication model proposed by Robson and Robinson considers the information provider as one major component in information seeking \cite{robson2013building}, drawing attention to their "needs, wants, and goals." In contrast, a common theme across previous projects is the predominant focus on the perspectives of information seekers, often sidelining challenges that providers might encounter \cite{seguin2022co}. For greater efficacy and longevity of relevant practice, it is vital to address these gaps and craft systems that accommodate needs of multiple stakeholders.

\section{Methods}
We conducted in-depth interviews with 21 participants in the United States (hereinafter referred to as the host country). Among them, 11 were international migrants (7 females, 4 males; average age = 26, S.E. = 1.69) with an average duration of 6.70 years of residency in the host country. The other 10 were domestic residents (5 females, 4 males, 1 non-binary; average age = 35, S.E. = 6.87). 

\subsection{Recruitment} 
We aimed to recruit migrant participants who have helped internationals from other home countries (referred as fellow migrants), and domestic residents with experience assisting migrants with daily information seeking. The recruitment took place through social media posts and physical flyers spread among our local community. The final research sample comprised 21 participants, varying in host country status, language background, and other demographic features.

\subsection{Data Collection}
We conducted semi-structured interviews with participants. Each interview lasted for about one hour. Migrant participants reflected on their life in the host country, detailed cases of assisting other migrants, and any support they had or wished to receive.  For participants who were domestic residents, we asked about their interactions with international migrants and their successes or failures in assisting them with daily information seeking. Interviews occurred face-to-face or via Zoom, based on the participant's choice. We reimbursed participants with \$15 U.S. dollar e-gift cards as a token of appreciation for their time. 

\section{Analysis}

We performed thematic analysis with textual data transcribed from the audio recordings \cite{braun2006using}. The analysis happened through an iterative process as new data was obtained. We stopped the analysis as well as the data collection when clear and stable themes identified. At the beginning, the leading author of this paper read through all interview transcripts at hand and developed an initial set of codes to cover all the ideas discussed by participants. In later steps, the entire research team examined this initial set of codes together, iterating between reviewing the existing interview data, obtaining new data, and revising the codes. By the end, we generated a codebook including 192 codes and 1370 quotations. It centered on themes that describe an information provider’s experience, including ways to establish connections with international migrants, information that they can offer to the migrants, challenges that prevent providers from issuing support, and reflections regarding possible ways to overcome the challenges.

\section{Findings}
Overall, the interviews suggested three types of challenges encountered by our migrant and domestic participants when offering informational support to others (RQ1) and their thoughts on future technology design (RQ2).  

\subsection{RQ1: Challenges Encountered by Information Providers}

\subsubsection{Language barriers and relevant social concerns.}
Participants reported the language barrier as their fundamental issues when providing their informational support. Both of our fellow migrant and domestic participants shared that they often failed to assist migrants with information seeking due to language barriers. For example, P1 shared his challenges of articulating his thoughts in front of other internationals:

\begin{quote}
“For a long time, I struggled to grasp the needs of other internationals. Many of us have accents. I didn’t know what they were talking about, and they couldn’t understand me either. We even tried typing texts during a face-to-face conversation to assist our communication in English.” [P1, fellow migrant]
\end{quote}

Additionally, many of our domestic participants reported that language barriers not only hindered the effective information exchange between migrants and domestic residents; it also resulted in negative social consequences undertaken by migrants, such as the worry about being devalued by native speakers and a sense of insecurity. P15 shared their experiences and observations in the above regard, which echoes similar points made by participants who assisted others as their fellow migrants (e.g., P3): 

\begin{quote}
    “Many international newcomers cannot speak or write English very well. My view is like, we all have those points, and we are not the most intelligent people out there. It’s common if you and I do not understand each other even among native speakers. However, I can tell they are hesitant to let me know that they can’t master the language. It is mostly because they lack confidence. They feel embarrassed.” [P15, domestic migrant]
\end{quote}

\begin{quote}
    “There are multiple levels of worries when I think of migrants seeking information from local people around. Of course, one of the worries is that people may not be able to articulate thoughts clearly. But there are also social concerns beyond that, like the worry that local people may view second-language speakers as outsiders and think they are not capable of doing stuff.” [P3, fellow migrant]
\end{quote}

Several domestic participants felt ill-equipped to bridge such social concerns of international migrants. They expressed that they "couldn't offer effective assistance if migrants remained hesitant to reach out." Notably, migrants in our sample never cited these social consequences when they spoke of interactions with other internationals, even though language barriers also existed in exchanges among migrants speaking different native languages. 

\subsubsection{Dynamic information landscapes and knowledge disparities.}

When participants reflected on the qualifications necessary for individuals to effectively assist others with their daily information seeking, they converged on one particular point: holding sufficient knowledge about the local systems, infrastructure, and services. Participants noted that while both fellow migrants and domestic residents can act as information providers, each group brings distinct benefits and confronts unique challenges when aiding migrant information seekers.

Migrant participants shared that to develop sufficient local knowledge, individuals usually need to stay long enough in the host country. Many gained the related knowledge through their daily experience or word of mouth in their social circle. Migrants in our sample felt that international migrants usually experienced common struggles while navigating local resources and services. The shared problems enabled participants to offer a unique piece of local knowledge that is helpful to other migrants:  

\begin{quote}
    “For all the internationals who study or work overseas, we might not be facing the exact same problem, but we face the same set of stressors. For example, most of us have trouble understanding the local healthcare system or buying a car. If one of us has learned something, this person can share the information with anybody who is not familiar with the system.” [P3, fellow migrant]
\end{quote}

Migrant participants noted they often provided informational support regarding knowledge specifically tied to their migration status, such as nuances related to visa regulations, the process of renewing a work permit, or the experience of transitioning from student to employment status in the host country. This distinct information, largely unavailable or unfamiliar to the domestic residents, can be helpful for new international migrants navigating similar paths. One example shared by our migrant participants:

\begin{quote}
    “We tend to know more about the rules and laws that apply to their visa for a temporary stay in this country. One example is optional practical training (OPT) required for us to work after graduation. It is specific to international students. Domestic people do not really understand that information or they may not care much, to be honest.” [P11, fellow migrant] 
\end{quote}

Meanwhile, migrant participants mentioned their issue of keeping up with the changing information landscape. While some migrant participants reflected on their experience of acting as providers for other migrants, they reported that staying up to date with the latest resources, services and practices could be challenging. This is particularly true for migration-related information needs. Participants acknowledged their importance but also recognized the intermittent and singular nature: some migration-related information needs are critical, but once the issue is resolved, it generally does not require attention again for several years. This infrequency poses challenges in keeping knowledge current and accurate. As explained by our participant: 

\begin{quote}
“The policies are changing fast. I knew another international student in the program. He asked me about travel regulations with our visa. I suggested scheduling an in-person appointment with his advisor to update his travel signature. However, it turned out my suggestion was not helpful. He sent couple of emails and found out it can be done online now. Many practices were updated during the pandemic, and obviously, neither me nor my friend could not keep track of all these information.” [P10, fellow migrant]
\end{quote}

Compared with migrants, one distinct benefit of domestic residents is that the latter have rich experiential knowledge about local systems and essential services in the host country. As described by our participants, this knowledge can help the former stay aware of available resources that they would not be able to discover otherwise: 

\begin{quote}
    “Domestic people know all the choices that can help. They also understand the pros and cons of each choice. However, migrants usually do not have much background knowledge [about the host country’s local situation], and it can be hard for them to choose among different information sources. So, domestic people can definitely help us to decide which one is more helpful.”  [P4, fellow migrant]
\end{quote}

\begin{quote}
    “A lot of the information and support would be more local, wouldn't it? Banking, employment opportunities, grocery stores... Local people know where they are and how to access them. That's where a local helper would be most beneficial. A newcomer can certainly learn the bus route and timetable by searching things online. But a local person will be able to tell them like, ‘this bus always runs 10 min late’ or ‘there's 2 different bus lines.’ Those detailed information can be hidden but they are very helpful.” [P21, domestic resident]
\end{quote}

In particular, domestic participants in our sample shared that it was sometimes difficult to comprehend the challenges of migrants. They made a note that there was often “a gap between what local people think the migrants know and what the migrants actually know.” Domestic participants often carried over their assumptions of international migrant groups when offering help. For instance, P14 and P16 worked as volunteers who assisted migrants in the local communities. They emphasized the importance of removing themselves from certain assumptions before communicating with migrants and learning their needs:

\begin{quote}
    “Many local people assume that, because migrants passed the TOEFL exam or something like that, they must be ready to go. You can just throw them into the world. That assumption is not true. You can't assume people have resources internally based on one standardized test. Also, local people often assume that migrants can use the local infrastructure that's in place, which is not necessarily true either.” [P16, domestic resident]
\end{quote}

\begin{quote}
    “It would be helpful if I knew what systems this migrant has in their home country and how they work. When a migrant comes to me for help, I often start with asking them questions like, ‘how would you seek a medical appointment in your home country?’ After they share, I will realize how their systems differ from ours [in the host country]. Knowing that difference will help me explain things clearly.” [P14, domestic resident]
\end{quote}

\subsubsection{Paradox between "feeling motivated" and "hesitating to commit."} 

Participants in our study acknowledged that fostering and maintaining the motivation for information providers to offer assistance is crucial, as the sustainability of these networks relies upon individuals’ willingness to help. Our analysis uncovered that the two groups of information providers had distinct motivations but faced similar challenges related to commitment.

Since fellow migrants hold migrant-oriented information that may benefit the daily practices of others, this commonality served more as their foundation to build emotional alignment with each other. Migrant participants reported a strong empathetic pull to aid others who are in similar situations. Their motivation to help is rooted in shared experiences and the desire to alleviate similar hardships for others.  As elaborated by our migrant participant:

\begin{quote}
    “I often grapple with my own issues – whether housing, or finding out the restrictions for job search. I have been in these situations and want to ease the journey for others. I know how important my help can be to those in the same boat.” [P1, fellow migrant]
\end{quote}

Our domestic participants also reflected on their motivations for assisting international migrants with their information-seeking needs. A recurring theme was the sense of personal fulfillment they experienced from assisting people in their new journey. They described the rewards of witnessing firsthand the impact of their support, as migrants became more confident and independent in their new environment. 

Acting as a helper for others often burdens a person’s already busy life. Migrant participants in our sample reported that much of their assistance for other migrants' information-seeking happened during their initial year of settlement. People spent most of their energy coping with acculturation stress. As a result, they found themselves lacking the bandwidth to commit to more responsibilities. As P10 noted:

\begin{quote}
    “For a lot of internationals, we have to try very hard to focus on academic performance, meet the requirements at work, and deal with homesickness. There is a lot to handle while we still struggle with listening and understanding everyone in English. We just do not have enough time to take care of anything else.” [P10, fellow migrant]
\end{quote}

Similarly, domestic participants expressed hesitation to engage in volunteer programs, especially those that require longer-term support and one-on-one mentorship for migrants. Participants shared that the commitment to continuous assistance can be substantial, requiring considerable time and energy. They voiced concerns about their capacity to sustain such intensive involvement in the face of their personal and professional obligations. Instead, participants desired some ways to contribute in a more flexible manner, as shared by our participant P17:

\begin{quote}
    “I spent half a year volunteering in a program matched with two mentees from China and Mexico. It was a rewarding experience, but sometimes overwhelming. My job [in the company] involves a lot of fieldwork, and can get pretty busy, leaving little room for my commitment [to supporting others] … I would need more flexible ways to get involved, like responding to requests whenever I am free. Just trying to support others while still maintaining balance in our personal lives.” [P17, domestic resident]
\end{quote}

The quote from P15 also highlights such challenges faced by many domestic residents wishing to assist international migrants. P15 served as a volunteer in a local English conversation group after his retirement. He shared that the long-term commitment associated with such volunteer work can be daunting for many potential providers, such as his family member, leading people to hesitate or opt-out of the program. This may exclude a large proportion of individuals who have the willingness and ability to contribute but are deterred by the perceived time commitment.

\begin{quote}
    “My wife saw the value and impact of what we do [as volunteers]. But she's concerned about over-committing as she isn't retired yet. Often, she would gather flyers in our neighborhoods to pass on me for distribution in my [volunteer] group. I mean, considering people like her, there's a vast potential support network out there; people would need to explore ways to tap into it without imposing overwhelming obligations.” [P15, domestic resident]
\end{quote}

\subsection{RQ2: ICT Design for the Interaction Between Information Providers and Seekers}

\subsubsection{Making deliberate use of language technologies.} 

Participants reflected on the pros and cons of leveraging AI-powered language technologies, such as machine translation, for the communication between information providers and seekers of different language backgrounds. Several domestic participants in our sample considered machine translation a promising technology to bridge the language barriers. They believed that AI-powered language tools have the potential to offer international migrants more independence, as they would “give migrants a sense of control on their second language." 

In contrast, the migrant participants were more aware of the technology’s limitations due to their prior experiences. For the latter group, it is essential that ICT platforms not only incorporate the machine translation module but also indicate the risk of counting on sometimes-imperfect translation output: 

\begin{quote}
    “Google Translate could generate translation outcomes that do not follow people’s natural way of talking. It could be weird to use their words or phrases when we provide information to people. It could let others form the wrong impression about you. For example, you may sound condescending when helping others because Google translates your message in a way that includes uncommon words and complex sentence structures.” [P7, fellow migrant]
\end{quote}

\begin{quote}
    “People can talk through the emotional aspect of their needs. However, machines may have a really difficult time understanding the emotion conveyed in a message and the severity of it. If I was using machines to translate information I offered, they [seekers] should pay more attention to the factual aspect of the request. Emotions should be handled by human translators.” [P8, fellow migrant]

\end{quote}

\subsubsection{Identifying shared knowledge for effective information exchange.}

Domestic participants reflected on their practices that aimed to contextualize the information seeker’s needs. They reported using shared vocabularies and knowledge as a starting point to fully comprehend the information needs of migrant seekers. Participants (P14 and P20) shared one example from when they served as volunteers at the city library to assist migrants with information seeking. When seekers’ information needs were ambiguous, volunteers tried to find common ground—words and concepts they both understood. Volunteers started with simple openers such as “Do you have a library card?” or “What do you need today”? to initiate the conversation with the information seeker. This approach, as P14 explained, is strategic: 

\begin{quote}
    “When they [migrant seekers] come in the library and say hi, the first question I will ask is, do you have a library card? What do you need today? Because that’s the starting point to know their needs. If I have problems understanding their requests, I will catch the keywords and try to figure out the shared vocabulary and knowledge between us. When I catch the keyword resume, I will ask, do you know how to use Indeed? Do you need to make a cover letter? Just ask questions using more common words so that people can easily answer with yes or no. Then I put these pieces together to figure out what people really need.” [P14, domestic resident]
\end{quote}

Asking “Do you have a library card?” is a tactical approach in this context. The library card symbolizes a shared understanding between the information provider and seeker. The question along with other yes/no questions, is part of a strategy to narrow down the seeker’s needs and guide the interaction toward a productive outcome. 

We communicated the above strategies to migrant participants in our sample for their feedback. They considered the practice helpful, but soon realized it would be inefficient and time-consuming for complex queries as “some of the requests might be broader than yes or no questions.” More importantly, continuous probing with questions might frustrate some migrant participants, who shared that being asked a lot of questions may make people “feel like being talked down to.” Participants suggested tools that allow information seekers and providers to establish a common ground for more effective assistance. For example, participant P17 shared: 

\begin{quote}
    “A cheatsheet outlining similarities and differences between systems in a migrant’s home country and here could be a great help. For example, while online banking is globally similar, paper-based banking like check deposits here confuses many migrants. A guide that provides a side-by-side comparison of systems, tailored to a person's nationality, could highlight the contrasts between two countries. This, in turn, would allow us to provide specifically adapted guidance based on their contexts.” [P17, domestic resident]
\end{quote}

\subsubsection{Enabling the communication of know-how.}

In the long term, migrants may benefit more from learning how to navigate the host country’s information system themselves than by relying on the help of others. To this end, several participants shared their thoughts regarding future ICT tools to enable teaching and learning know-how that promote self-reliance and long-term success. Our participants emphasized the importance of equipping migrants with the skills and knowledge to become independent, rather than just offering timely assistance. Such insights revolved around generating user-tailored tutorials as the core feature of the design. 

\begin{quote}
    “If a migrant comes to me for help, I am always happy to give them the information they need. However, it is more important to scaffold them for the future. For example, I can screen capture what I do, and then share it with the person in a file. Because that way they can watch what it is and then be like, ‘oh that was easy’. It is not really easy until you have seen the whole process done. In the future, maybe the system can automatically generate a video for the person to follow. They will get the right information by following the video.” [P13, domestic resident]
\end{quote}

\section{Discussion}

By incorporating the perspectives of information providers, we share results that highlight the barriers that hinder them from providing effective assistance to migrants.  The current research deepened our understanding of the information practices of international migrants by focusing on their information providers, an often-overlooked role. Our study highlights the unique benefits each group of information providers brings to migrants' information-seeking practices. For instance, domestic residents can offer nuanced information about local communities, while fellow migrants can guide migration-related topics. These findings are consistent with prior research (e.g., \cite{komito2011migrants,mao2015investigating}), and fit well within existing information-seeking frameworks that discuss information sources' characteristics and utility. Furthermore, our interviews highlighted the challenges faced by these information providers. Fellow migrants, for example, occasionally struggled to stay updated of the ever-evolving migration policies and regulations, while domestics sometimes found it hard to contextualize migrants' inquiries. Existing models within LIS focus primarily on information seeking and the user (e.g., \cite{ellis1989behavioural,johnson1997cancer,wilson1981user}), often offering a simplified view of information sources. However, the challenges relayed by our participants underscore the intricate and multifaceted nature of understanding international migrants' information-seeking experiences. We recommend that future scholars factor in the perspectives of international migrants' potential providers when assessing their information-seeking practices and when designing related technologies.

Furthermore, our research revealed significant knowledge gaps between international migrants and a key group of information providers, domestic residents who often struggle to understand the context behind a migrant's request and recognize their unique challenges. Previous work, such as those of \cite{brown2016designing,kukulska2015mobile,schwarz2015help}, mostly emphasized the value of local knowledge in supporting newcomer information-seeking. Specifically, Brown and Grinter developed a messaging platform facilitating communication between resettling refugees and American families \cite{brown2016designing}. Over 12 weeks, they exchanged messages on diverse daily topics. This project showcased the potential of domestic residents using their local insights to assist migrants. While such local knowledge is vital, our findings suggest that not understanding migrants' home country systems can hinder effective communication. This limited understanding demands more effort to establish common ground, potentially degrading communication quality \cite{wang2009cultural,yuan2013understanding}, and information-seeking outcomes \cite{shah2015collaborative}. Most existing designs emphasize assisting migrants with learning about local knowledge. Our results hint at the opportunity for information providers to grasp more about migrants' backgrounds, which, consequently, may lessen the stress undertaken by the migrants \cite{ward2001psychology}. We suggest future ICTs incorporate features that share relevant contextual information about migrants with domestic providers.

In addition, participants in our study also reflected on their commitment and concerns about their balance with life. They expressed a desire to participate in the activities of assisting migrants with information seeking in a more flexible manner. That is, participants suggested that they could possibly be involved in task-based commitment instead of longer-term commitment (e.g., paired mentorship). Failures to provide consistent support may negatively affect migrant seekers’ user experience, thus reducing their willingness to reach out, and trust towards these programs. For example, a two-month project was launched in Austria and the United Kingdom to pair immigrants with nearby registered volunteers to provide daily essential information \cite{schwarz2015help}. One of the two Arabic-speaking volunteers dropped out in the fifth week, resulting in frequent unavailability of information providers. The Arabic immigrant group generated significantly less interaction and usage of the service by the end of the program. We suggested that future systems consider providing more alternatives for participation. For example, it can follow the design of an online freelance marketplace (e.g., TaskRabbit) to offer greater flexibility to people depending on their availability \cite{hoque_how_2015}.

\section{Limitations}

Our study has several limitations. Despite efforts for diverse recruitment, our sample may not represent the full range of potential information providers beyond a migrant's co-nationals. We primarily used online platforms, possibly excluding those with limited tech literacy. While our sample included information providers who were migrants themselves, these participants all have lived in the host country for less than 10 years. Compared with other individuals qualified for our definition of fellow migrants, these participants may share unique commonalities in assisting another migrant's daily information seeking. The transferability of our findings should consider these distinctions.

\section{Conclusion}
Through in-depth interviews with 21 information providers, including fellow migrants coming from different home countries and domestic residents, we identified challenges that hindered people from offering information support to international migrants. These challenges consider how information providers deal with language barriers, overcome knowledge disparities, and calibrate their effort commitment when assisting with a migrant's informational needs. Our work shed light on design directions for future ICT that facilitate international migrants' daily information seeking by accounting for the provider's needs and concerns.

\bibliographystyle{splncs04}
\bibliography{finalpaper}

\begin{thebibliography}{10}
\providecommand{\url}[1]{\texttt{#1}}
\providecommand{\urlprefix}{URL }
\providecommand{\doi}[1]{https://doi.org/#1}

\bibitem{alberts2005there}
Alberts, H.C., Hazen, H.D.: “there are always two voices…”: International
  students' intentions to stay in the united states or return to their home
  countries. International migration  \textbf{43}(3),  131--154 (2005)

\bibitem{alho2020you}
Alho, R.: ‘you need to know someone who knows someone’: international
  students’ job-search experiences. Nordic journal of working life studies
  (2020)

\bibitem{alzougool2013finding}
Alzougool, B., Chang, S., Gomes, C., Berry, M.: Finding their way around:
  International students’ use of information sources. Journal of Advanced
  Management Science  \textbf{1}(1),  43--49 (2013)

\bibitem{berendt2009user}
Berendt, B., Kralisch, A.: A user-centric approach to identifying best
  deployment strategies for language tools: the impact of content and access
  language on web user behaviour and attitudes. Information Retrieval
  \textbf{12},  380--399 (2009)

\bibitem{blasi2021linguistic}
Blasi, D.E., Mishra, V., Garc{\'\i}a, A.M., Dexter, J.P.: Linguistic fairness
  in the us: The case of multilingual public health information about covid-19.
  medRxiv pp. 2021--09 (2021)

\bibitem{bochner1977friendship}
Bochner, S., McLeod, B.M., Lin, A.: Friendship patterns of overseas students: A
  functional model 1. International journal of psychology  \textbf{12}(4),
  277--294 (1977)

\bibitem{braun2006using}
Braun, V., Clarke, V.: Using thematic analysis in psychology. Qualitative
  research in psychology  \textbf{3}(2),  77--101 (2006)

\bibitem{brown2016designing}
Brown, D., Grinter, R.E.: Designing for transient use: A human-in-the-loop
  translation platform for refugees. In: Proceedings of the 2016 CHI conference
  on human factors in computing systems. pp. 321--330 (2016)

\bibitem{caidi2008information}
Caidi, N., Allard, D., Dechief, D.: Information practices of immigrants to
  canada: A review of the literature. Unpublished report to Citizenship and
  Immigration Canada  (2008)

\bibitem{caidi2010information}
Caidi, N., Allard, D., Quirke, L.: Information practices of immigrants. Annual
  review of information science and technology  \textbf{44}(1),  491--531
  (2010)

\bibitem{case2010model}
Case, D.O.: A model of the information seeking and decision making of online
  coin buyers. Information Research  \textbf{15}(4),  15--4 (2010)

\bibitem{chang2017digital}
Chang, S., Gomes, C.: Digital journeys: A perspective on understanding the
  digital experiences of international students. Journal of International
  Students  \textbf{7}(2),  347--466 (2017)

\bibitem{chang2020way}
Chang, S., McKay, D., Caidi, N., Mendoza, A., Gomes, C., Ekmekcioglu, C.: From
  way across the sea: Information overload and international students during
  the covid-19 pandemic. Proceedings of the Association for Information Science
  and Technology  \textbf{57}(1), ~e289 (2020)

\bibitem{chatman1991life}
Chatman, E.A.: Life in a small world: Applicability of gratification theory to
  information-seeking behavior. Journal of the American Society for information
  science  \textbf{42}(6),  438--449 (1991)

\bibitem{chung2015exploratory}
Chung, E., Yoon, J.: An exploratory analysis of international students'
  information needs and uses/exploration et analyse des besoins et des
  utilisations d'information des {\'e}tudiants internationaux. Canadian Journal
  of Information and Library Science  \textbf{39}(1),  36--59 (2015)

\bibitem{cobo2013exploration}
Cobo, C.: Exploration of open educational resources in non-english speaking
  communities. International Review of Research in Open and Distributed
  Learning  \textbf{14}(2),  106--128 (2013)

\bibitem{ellis1989behavioural}
Ellis, D.: A behavioural approach to information retrieval system design.
  Journal of documentation  \textbf{45}(3),  171--212 (1989)

\bibitem{forbush2016social}
Forbush, E., Foucault-Welles, B.: Social media use and adaptation among chinese
  students beginning to study in the united states. International Journal of
  Intercultural Relations  \textbf{50},  1--12 (2016)

\bibitem{gao2022taking}
Gao, G., Zheng, J., Choe, E.K., Yamashita, N.: Taking a language detour: How
  international migrants speaking a minority language seek covid-related
  information in their host countries. Proceedings of the ACM on Human-Computer
  Interaction  \textbf{6}(CSCW2),  1--32 (2022)

\bibitem{gatteschi2012match}
Gatteschi, V., Lamberti, F., Demartini, C.: Lo-match: A semantic platform for
  matching migrants' competences with labour market's needs. In: Proceedings of
  the 2012 IEEE Global Engineering Education Conference (EDUCON). pp.~1--5.
  IEEE (2012)

\bibitem{han2015mental}
Han, M., Pong, H.: Mental health help-seeking behaviors among asian american
  community college students: The effect of stigma, cultural barriers, and
  acculturation. Journal of College Student Development  \textbf{56}(1),  1--14
  (2015)

\bibitem{hendrickson2011analysis}
Hendrickson, B., Rosen, D., Aune, R.K.: An analysis of friendship networks,
  social connectedness, homesickness, and satisfaction levels of international
  students. International journal of intercultural relations  \textbf{35}(3),
  281--295 (2011)

\bibitem{hirsch2004speakeasy}
Hirsch, T., Liu, J.: Speakeasy: overcoming barriers and promoting community
  development in an immigrant neighborhood. In: Proceedings of the 5th
  conference on Designing interactive systems: processes, practices, methods,
  and techniques. pp. 345--348 (2004)

\bibitem{hoque_how_2015}
Hoque, F.: How {The} {Rising} {Gig} {Economy} {Is} {Reshaping} {Businesses}
  (Sep 2015),
  \url{https://www.fastcompany.com/3051315/the-gig-economy-is-going-global-heres-why-and-what-it-means}

\bibitem{hurh1990religious}
Hurh, W.M., Kim, K.C.: Religious participation of korean immigrants in the
  united states. Journal for the scientific study of religion pp. 19--34 (1990)

\bibitem{jackson2005incoming}
Jackson, P.A.: Incoming international students and the library: a survey.
  Reference services review  \textbf{33}(2),  197--209 (2005)

\bibitem{jeong2004unbreakable}
Jeong, W.: Unbreakable ethnic bonds: Information-seeking behavior of korean
  graduate students in the united states. Library \& Information Science
  Research  \textbf{26}(3),  384--400 (2004)

\bibitem{johnson1997cancer}
Johnson, J.D.: Cancer-related information seeking. Hampton Press, Creskill, NJ.
   (1997)

\bibitem{johnson2012health}
Johnson, J.D., Case, D.O.: Health information seeking, vol.~52. Peter Lang New
  York, NY (2012)

\bibitem{komito2011migrants}
Komito, L., Bates, J.: Migrants' information practices and use of social media
  in ireland: Networks and community. p. 289–295. iConference '11,
  Association for Computing Machinery, New York, NY, USA (2011).
  \doi{10.1145/1940761.1940801}, \url{https://doi.org/10.1145/1940761.1940801}

\bibitem{kudo2003intercultural}
Kudo, K., Simkin, K.A.: Intercultural friendship formation: The case of
  japanese students at an australian university. Journal of intercultural
  studies  \textbf{24}(2),  91--114 (2003)

\bibitem{kuhlthau1991inside}
Kuhlthau, C.C.: Inside the search process: Information seeking from the user's
  perspective. Journal of the American society for information science
  \textbf{42}(5),  361--371 (1991)

\bibitem{kukulska2015mobile}
Kukulska-Hulme, A., Gaved, M., Paletta, L., Scanlon, E., Jones, A., Brasher,
  A.: Mobile incidental learning to support the inclusion of recent immigrants.
  Ubiquitous Learning: an international journal  \textbf{7}(2),  9--21 (2015)

\bibitem{lee2011internet}
Lee, E.J., Lee, L., Jang, J.: Internet for the internationals: Effects of
  internet use motivations on international students' college adjustment.
  Cyberpsychology, Behavior, and Social Networking  \textbf{14}(7-8),  433--437
  (2011)

\bibitem{liao2007information}
Liao, Y., Finn, M., Lu, J.: Information-seeking behavior of international
  graduate students vs. american graduate students: A user study at virginia
  tech 2005. College \& Research Libraries  \textbf{68}(1),  5--25 (2007)

\bibitem{liu2009chinese}
Liu, G., Winn, D.: Chinese graduate students and the canadian academic library:
  A user study at the university of windsor. The Journal of Academic
  Librarianship  \textbf{35}(6),  565--573 (2009)

\bibitem{liu1997information}
Liu, M., Redfern, B.: Information-seeking behavior of multicultural students: a
  case study at san jose state university. College \& Research Libraries
  \textbf{58}(4),  348--354 (1997)

\bibitem{mao2015investigating}
Mao, Y.: Investigating chinese migrants' information-seeking patterns in
  canada: media selection and language preference. Global Media Journal
  \textbf{8}(2), ~113 (2015)

\bibitem{maundeni2001role}
Maundeni, T.: The role of social networks in the adjustment of african students
  to british society: students' perceptions. Race Ethnicity and Education
  \textbf{4}(3),  253--276 (2001)

\bibitem{mehra2007glocal}
Mehra, B., Papajohn, D.: “glocal” patterns of communication-information
  convergences in internet use: Cross-cultural behavior of international
  teaching assistants in a culturally alien information environment. The
  International Information \& Library Review  \textbf{39}(1),  12--30 (2007)

\bibitem{ngan2016refugees}
Ngan, H.Y., Lifanova, A., Jarke, J., Broer, J.: Refugees welcome: Supporting
  informal language learning and integration with a gamified mobile
  application. In: Adaptive and Adaptable Learning: 11th European Conference on
  Technology Enhanced Learning, EC-TEL 2016, Lyon, France, September 13-16,
  2016, Proceedings 11. pp. 521--524. Springer (2016)

\bibitem{oh2016newcomers}
Oh, C.Y., Butler, B.S.: Newcomers from the other side of the globe:
  International students' local information seeking during adjustment.
  Proceedings of the Association for Information Science and Technology
  \textbf{53}(1), ~1--6 (2016)

\bibitem{oh2014information}
Oh, C.Y., Butler, B.S., Lee, M.: Information behavior of international students
  settling in an unfamiliar geo-spatial environment. Proceedings of the
  American Society for Information Science and Technology  \textbf{51}(1),
  1--11 (2014)

\bibitem{pavlenko2008m}
Pavlenko, A.: “i'm very not about the law part”: Nonnative speakers of
  english and the miranda warnings. Tesol Quarterly  \textbf{42}(1),  1--30
  (2008)

\bibitem{robson2013building}
Robson, A., Robinson, L.: Building on models of information behaviour: linking
  information seeking and communication. Journal of documentation
  \textbf{69}(2),  169--193 (2013)

\bibitem{rozsa2015online}
R{\'o}zsa, G., Komlodi, A., Chu, P.: Online searching in english as a foreign
  language. In: Proceedings of the 24th International Conference on World Wide
  Web. pp. 875--880 (2015)

\bibitem{rui2015social}
Rui, J.R., Wang, H.: Social network sites and international students’
  cross-cultural adaptation. Computers in Human Behavior  \textbf{49},
  400--411 (2015)

\bibitem{sakurai2010building}
Sakurai, T., McCall-Wolf, F., Kashima, E.S.: Building intercultural links: The
  impact of a multicultural intervention programme on social ties of
  international students in australia. International journal of intercultural
  relations  \textbf{34}(2),  176--185 (2010)

\bibitem{savolainen2005everyday}
Savolainen, R.: Everyday life information seeking: Approaching information
  seeking in the context of “way of life”. Library \& information science
  research  \textbf{17}(3),  259--294 (1995)

\bibitem{schwarz2015help}
Schwarz, S., Salazar, E.P., Bobeth, J., Bersia, N., Tscheligi, M.: Help radar:
  ubiquitous assistance for newly arrived immigrants. In: Proceedings of the
  14th International Conference on Mobile and Ubiquitous Multimedia. pp.
  183--194 (2015)

\bibitem{seguin2022co}
Seguin, J.P., Varghese, D., Anwar, M., Bartindale, T., Olivier, P.:
  Co-designing digital platforms for volunteer-led migrant community welfare
  support. In: Designing Interactive Systems Conference. pp. 247--262 (2022)

\bibitem{shah2015collaborative}
Shah, C.: Collaborative information seeking: From ‘what?’and ‘why?’to
  ‘how?’and ‘so what?’. Collaborative information seeking: best
  practices, new domains and new thoughts pp. 3--16 (2015)

\bibitem{sin2015demographic}
Sin, S.C.J.: Demographic differences in international students' information
  source uses and everyday information seeking challenges. The Journal of
  Academic Librarianship  \textbf{41}(4),  466--474 (2015)

\bibitem{sin2011international}
Sin, S.C.J., Kim, K.S., Yang, J., Park, J.A., Laugheed, Z.T.: International
  students' acculturation information seeking: Personality, information needs
  and uses. Proceedings of the American Society for Information Science and
  Technology  \textbf{48}(1), ~1--4 (2011)

\bibitem{smith2011review}
Smith, R.A., Khawaja, N.G.: A review of the acculturation experiences of
  international students. International Journal of intercultural relations
  \textbf{35}(6),  699--713 (2011)

\bibitem{suh2019had}
Suh, M., Hsieh, G.: The “had mores”: Exploring korean immigrants’
  information behavior and ict usage when settling in the united states.
  Journal of the Association for Information Science and Technology
  \textbf{70}(1),  38--48 (2019)

\bibitem{tang2018new}
Tang, C., Gui, X., Chen, Y., Magueramane, M.: New to a country: Barriers for
  international students to access health services and opportunities for
  design. In: Proceedings of the 12th EAI International Conference on Pervasive
  Computing Technologies for Healthcare. pp. 45--54 (2018)

\bibitem{thomas2010transnational}
Thomas, F.: Transnational health and treatment networks: meaning, value and
  place in health seeking amongst southern african migrants in london. Health
  \& Place  \textbf{16}(3),  606--612 (2010)

\bibitem{tsai2021help}
Tsai, C.H., Gui, X., Kou, Y., Carroll, J.M.: With help from afar: Cross-local
  communication in an online covid-19 pandemic community. Proceedings of the
  ACM on Human-Computer Interaction  \textbf{5}(CSCW2),  1--24 (2021)

\bibitem{wang2009cultural}
Wang, H.C., Fussell, S.F., Setlock, L.D.: Cultural difference and adaptation of
  communication styles in computer-mediated group brainstorming. In:
  Proceedings of the SIGCHI conference on human factors in computing systems.
  pp. 669--678 (2009)

\bibitem{wang2015immigration}
Wang, L., Kwak, M.J.: Immigration, barriers to healthcare and transnational
  ties: A case study of south korean immigrants in toronto, canada. Social
  Science \& Medicine  \textbf{133},  340--348 (2015)

\bibitem{ward2001psychology}
Ward, C., Bochner, S., Furnham, A.: The {Psychology} of {Culture} {Shock} (Jan
  2001). \doi{10.4324/9781003070696}

\bibitem{ward1991impact}
Ward, C., Searle, W.: The impact of value discrepancies and cultural identity
  on psychological and sociocultural adjustment of sojourners. International
  journal of intercultural relations  \textbf{15}(2),  209--224 (1991)

\bibitem{williams2011can}
Williams, C.T., Johnson, L.R.: Why can’t we be friends?: Multicultural
  attitudes and friendships with international students. International Journal
  of Intercultural Relations  \textbf{35}(1),  41--48 (2011)

\bibitem{wilson1981user}
Wilson, T.D.: On user studies and information needs. Journal of documentation
  \textbf{37}(1),  3--15 (1981)

\bibitem{wilson1994information}
Wilson, T.D.: Information needs and uses: fifty years of progress. Fifty years
  of information progress: a Journal of Documentation review  \textbf{28}(1),
  15--51 (1994)

\bibitem{wyche2012we}
Wyche, S.P., Grinter, R.E.: " this is how we do it in my country" a study of
  computer-mediated family communication among kenyan migrants in the united
  states. In: Proceedings of the ACM 2012 conference on Computer Supported
  Cooperative Work. pp. 87--96 (2012)

\bibitem{yan2013chinese}
Yan, K., Berliner, D.C.: Chinese international students' personal and
  sociocultural stressors in the united states. Journal of college student
  development  \textbf{54}(1),  62--84 (2013)

\bibitem{ye2006traditional}
Ye, J.: Traditional and online support networks in the cross-cultural
  adaptation of chinese international students in the united states. Journal of
  Computer-Mediated Communication  \textbf{11}(3),  863--876 (2006)

\bibitem{yeh2003international}
Yeh, C.J., Inose, M.: International students' reported english fluency, social
  support satisfaction, and social connectedness as predictors of acculturative
  stress. Counselling Psychology Quarterly  \textbf{16}(1),  15--28 (2003)

\bibitem{yi2007international}
Yi, Z.: International student perceptions of information needs and use. The
  Journal of Academic Librarianship  \textbf{33}(6),  666--673 (2007)

\bibitem{yoon2017international}
Yoon, J., Chung, E.: International students’ information needs and seeking
  behaviours throughout the settlement stages. Libri  \textbf{67}(2),  119--128
  (2017)

\bibitem{yuan2013understanding}
Yuan, C.W., Setlock, L.D., Cosley, D., Fussell, S.R.: Understanding informal
  communication in multilingual contexts. In: Proceedings of the 2013
  conference on Computer supported cooperative work. pp. 909--922 (2013)

\end{thebibliography}
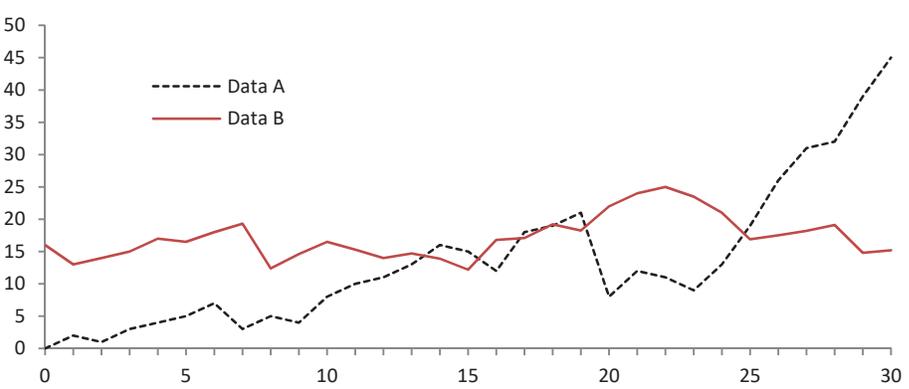
\end{document}